# Solving Boolean Satisfiability Problems Using A Hypergraph-based Probabilistic Computer

Yihan He[1†], Ming-Chun Hong[2,3†],Wanli Zheng[1],Ching Shih[2,3], Hsin-Han Lee[3], Yu-Chen Hsin[3], Jeng-Hua Wei[3], Xiao Gong[1], Tuo-Hung Hou[2,4], and Gengchiau Liang[1,4*]

[1]Department of Electrical and Computer Engineering, National University of Singapore, 117583 Singapore
[2]Department of Electrical Engineering and Institute of Electronics, National Yang-Ming Chiao Tung University, Hsinchu, Taiwan
[3]Electronic and Optoelectronic System Research Laboratories, Industrial Technology Research Institute, Hsinchu, Taiwan
[4]Industry Academia Innovation School, National Yang-Ming Chiao Tung University, Hsinchu, Taiwan

[†]These authors contribute equally; [*]Email: gcliang@nycu.edu.tw

## Abstract

***Boolean Satisfiability (SAT) problems are critical in fields such as artificial intelligence and cryptography, where efficient solutions are essential. Conventional probabilistic solvers often encounter scalability issues due to complex logic synthesis steps. In this work, we present a novel approach for solving the 3-SAT Boolean satisfiability problem using hypergraph-based probabilistic computers obtained through direct mapping. This method directly translates 3-SAT logical expressions into hypergraph structures, thereby circumventing conventional logic decomposition and synthesis procedures, and offering a more streamlined solver architecture. For representative uf100-430 instances, the proposed approach reduces the node count from 631 to 100 and the edge count from ~2,423 to ~1,013. Under identical simulated annealing conditions, the conventional simple undirected graph (SUG)-based solver achieves a 0% success rate across the tested instances, whereas the hypergraph-based solver attains an average success rate of ~77.6%. In addition, the hypergraph-based method reaches an average minimum energy of ~0.24, close to the theoretical ground state, while the SUG-based architecture remains trapped at substantially higher energy levels (~9.12 on average). The direct hypergraph mapping can further be extended to k-SAT formulations, providing a scalable framework for more complex satisfiability problems in probabilistic computing.***

## Introduction

In recent years, probabilistic computing based on probabilistic bit (p-bit)[1–7] has been recognized as an unconventional computing paradigm in response to the diminishing returns of classical deterministic computing, particularly as the benefits of Moore's Law plateau. By leveraging the inherent randomness in physical systems, probabilistic computing enables computations based on probability distributions rather than fixed binary states. A particularly promising application of probabilistic computing is combinatorial optimization problems (COPs)[8–12], where classical computers struggle with scalability and efficiency. Based on the architecture of simple undirected graphs (SUGs), probabilistic computers constructed from Boltzmann machines (BMs)[10,13–15] or the Ising model[3,16–18] demonstrate superior capabilities in solving hard computational problems with vast solution space. Among these, the Boolean Satisfiability (SAT) problem[19], particularly the 3-SAT variant, stands out due to its wide applications in artificial intelligence[20], electronic design automation[21], and cryptography[22]. The 3-SAT problem seeks to determine a variable assignment that satisfies a Boolean formula composed of clauses, each containing exactly three literals.

Current probabilistic solvers based on invertible logic for SAT problems often rely on logic synthesis methodologies[23–25], as shown in **Fig. 1(a)**. These approaches initially construct small SUGs capable of implementing basic logical functions from the truth tables[26] or logical expressions of target logic gates[27]. Common examples include invertible NOT (INOT), invertible OR (IOR), and invertible AND (IAND) gates. Then these independent SUGs can be combined according to pre-designed logic diagrams to form larger-scale combinatorial invertible logic circuits to solve SAT problems. However, this approach faces great challenges in scalability. In a 3-SAT problem, each clause consists of three literals (variables or their negations), which are typically implemented in the graph by merging the SUGs of INOT and IOR gates, or by cascading two IOR gates. This process introduces an auxiliary node at each merging step. As the scale of 3-SAT problems increases, the growing number of clauses introduces a substantial number of similar auxiliary nodes. This proliferation of auxiliary nodes significantly increases the overall node overhead, exacerbates the structural complexity of the graph by enhancing connectivity and depth, and complicates the energy landscape of the system. Additionally, the exponential growth of the solution space places considerable demands on the solver's computational capabilities and hardware resources. Consequently, logic synthesis-based solvers encounter severe performance bottlenecks when addressing large-scale 3-SAT problems.

To address these limitations, recent research has explored the incorporation of higher-order or many-body interactions among nodes[16,28–30]. Studies have demonstrated that higher-order bipolar Ising machines, implemented with coupled oscillators[16,31], can find solutions more efficiently with fewer spin variables, i.e. less node overhead, compared to conventional second-order Ising machines. However, a critical limitation of this scheme is the deterministic representation of spin variables. It impedes the system's ability to escape local energy minimum points once trapped, thereby compromising the quality of solution and overall accuracy. As a result, a careful and appropriate design of annealing schedules is indispensable for ensuring convergence to the global optimum[32]. A reconfigurable CMOS Ising machine with direct mapping has also been proposed[33] and demonstrated the feasibility of mapping SAT clauses without intermediate decomposition. Nevertheless, its validation was limited to small-scale 3-SAT examples (e.g., illustrative 8-variable instances), which leaves scalability to larger and structurally more complex benchmarks largely unexplored. Motivated by these observations, we present a novel probabilistic computing approach based on invertible logic developed from BMs. Employing binary variable representation, the proposed design methodology can directly map 3-SAT logical expressions onto hypergraph structures (**Fig. 1b**), rather than merging SUGs. This design cleverly circumvents the intrinsic logic decomposition and synthesis procedures embedded in conventional probabilistic SAT solvers. Such a direct mapping technique offers three key advantages: (1) minimizing node overhead, (2) simplifying the structural complexity of the solver, and (3) achieving the theoretically minimal solution space. Meanwhile, the core of our probabilistic solver lies the Boltzmann distribution (**Fig. 1c**), which governs the stochastic sampling of state configurations according to their energy. To further improve the solution quality, we also incorporate simulated annealing by dynamically increasing the system's stochasticity level parameter to facilitate the system to converge toward the global optimum (ground state). By leveraging the natural randomness of p-bit devices and the annealing-enhanced Boltzmann sampling process, our solver demonstrates significantly improved success rates and computational efficiency on practical 3-SAT instances.

## Results

### Direct mapping 3-SAT to hypergraphs

The core of solving 3-SAT problems based on probabilistic computing resides in the formulation of an appropriate energy function that accurately represents the objective function or logical expression. This process must ensure that the optimal solution, which may correspond to one or multiple variable assignments, of the problem to be solved is mapped to the ground state of the system's energy landscape. The universally applicable mapping methodology we developed allows the direct construction of invertible logic-based probabilistic solvers for 3-SAT utilizing energy models based on hypergraph structures. This approach is also designed to be scalable, which can be extended to higher-order *k*-SAT problems, thus offering a versatile framework for broader applicability in complex SAT problem domains.

The formula of a 3-SAT problem is expressed as a conjunction (logical AND) of *m* clauses, each containing exactly three literals (variables or their negations) connected by disjunctions (logical OR). Formally, the general form of the 3-SAT problem can be represented as follows:

$$(x_1 \vee \neg x_2 \vee x_{N-3}) \wedge (\neg x_2 \vee x_4 \vee \neg x_{N-1}) \wedge \ldots \wedge (x_{N-2} \vee \neg x_6 \vee \neg x_N) \tag{1}$$

where $N$ denotes the total number of variables.

The objective of the 3-SAT problem is to find a combination of true or false assignments for the variables $x_1, x_2, \ldots, x_N$, that makes the entire formula true, which is a canonical NP-complete problem. To address the problem using our approach, we begin by defining an encoding rule that uniformly represents both positive and negative literals in a binary format. Specifically, the encoding of literals is given by:

$$x_i \rightarrow (1 - x_i) \tag{2}$$

$$\neg x_i \rightarrow x_i. \tag{3}$$

This encoding allows us to equally handle the literals within each clause. Applying the encoding from Eq. (2) and Eq. (3), we can derive individual energy functions for each clause:

$$\begin{gathered} H_1 = (1 - x_1) \cdot x_2 \cdot (1 - x_{N-3}) \\ H_2 = x_2 \cdot (1 - x_4) \cdot x_{N-1} \\ \vdots \\ H_m = (1 - x_{N-2}) \cdot x_6 \cdot x_N \end{gathered} \tag{4}$$

where each energy function $H_i$ represents a product of terms corresponding to the encoded literals. This reflects the constraints imposed by the disjunctions within each clause. The overall energy landscape of the 3-SAT problem can be obtained by summing the energy functions of all *m* clauses. Therefore, the total energy function $H_{total}$ for the entire 3-SAT problem is expressed as:

$$H_{total} = \sum_{i=1}^{m} H_i = \sum_{i<j<k} W_{ijk} \cdot x_i x_j x_k + \sum_{i<j} J_{ij} \cdot x_i x_j + \sum_{i} h_i \cdot x_i + \cdots + C \quad (5)$$

where $W_{ijk}$, $J_{ij}$ and $h_i$ represent the coefficients for hyperedges, pairwise interactions, and biases respectively, and $C$ is a constant term. These parameters directly correspond to the coefficients of the hypergraph.

The summation operation of each energy function encapsulates the collective constraints imposed by the conjunction of all clauses. Using this approach, the system reaches its ground state, namely $H_{total} = 0$, only when the variables $x_1, x_2, \ldots, x_N$ are assigned in such a way that the entire formula is satisfied, i.e., evaluates to true. Any other assignments of variables that do not satisfy the formula will return a total energy $H_{total}$ greater than zero.

Following the outlined mapping procedure, **Fig. 2(a)** presents an example of a 3-SAT problem consisting of 5 literals and 3 clauses. In its direct mapping to a hypergraph depicted in **Fig. 2(b)**, each variable is represented as one node, while the constraints between clauses and the relationships between literals within a clause are fully captured by hyperedges, pairwise interactions, and bias terms.

### Conventional SUGs-based logic synthesis

We also constructed a conventional 3-SAT solver using SUGs obtained through standard logic synthesis using INOT and IOR gates[25,34]. In this framework, negated literals are realized using INOT gates, while clause-level logical OR operations are implemented using IOR gates. During the synthesis process, the output node of each INOT gate must be merged with the corresponding input node of an IOR gate to represent negated variables within a clause. This merging operation inherently introduces auxiliary nodes.

**Fig. 2(c)** presents the matrix representations of the IOR and INOT SUGs used in this work. The pairwise interaction matrix $J_{ij}$ and the bias vector $h_i$ explicitly encode the energy formulation corresponding to each logic gate. The matrix elements quantify the interaction strength between node pairs. During logic synthesis, these gate-level SUGs are systematically merged to form the complete solver graph for the given 3-SAT instance. The right panel illustrates the resulting 12-node network for a representative small-scale 3-SAT example comprising five literals and three clauses. This synthesis strategy follows a standard and widely adopted invertible logic construction, the cascading of INOT and IOR gates inevitably increases the number of nodes and pairwise edges in the final graph. This structural overhead becomes more pronounced as the number of clauses increases, but it provides a meaningful baseline for comparison with the proposed hypergraph-based direct mapping approach.

### Tunable p-bit devices and systems

The implementation of p-bit device is various, including but not limited to magnets-based[4,35,36], memristors[8,37,38], FPGA[14,14,39], and ferroelectric transistors[40]. In this work, we adopted a stochastic spin-transfer-torque magnetic random access memory (STT-MRAM)[41] as a tunable p-bit device. The cross-sectional TEM image in **Fig. 3(a)** shows its layered composition. This device demonstrates distinctive switching characteristics across various voltage ranges: at a low applied voltage, the device stabilizes in the "*AP* state" (antiparallel state), whereas high voltage conditions drive it into the "$P$ state" (parallel state)[41]. Within an intermediate voltage range, the device enters a tunable oscillation regime. As shown in **Fig. 3(b)**, it fluctuates between the *"P"* and *"AP"* states. This voltage-dependent oscillation under medium voltage conditions is central to the device's operation as a tunable p-bit device. The stochastic behavior of the device is experimentally characterized in **Fig. 3(c)**, where the observed data (black squares) conform closely to a sigmoidal fitting curve (blue line), illustrating a smooth probabilistic transition between states in response to varying input voltage. This sigmoidal response is essential for achieving the tunable probabilistic output required for the probabilistic computing applications considered in this study. **Fig. 3(d)** shows the basic circuit configuration of the STT-biased probabilistic device. The temporal behavior of the device current under different bias voltages is further shown in **Fig. 3(e)**. In our further modeling and system-level simulations in solving the 3-SAT problems, we extracted this sigmoidal curve and encapsulated it into a cell with a probability of 50% to get 0 or 1 at $V = 0$. Moreover, to represent the involved three-order interactions of binary states in hypergraphs, we employed standard CMOS AND gates, as shown in **Fig. 3(f)**. It should be noted that, in the hardware realization, a proper interface stage needs to be inserted between the p-bit and the digital logic. Specifically, the bipolar p-bit output (e.g., $\pm V_{DD}/2$) first needs to be processed by a sense amplifier or comparator to convert the bipolar signal into a CMOS-compatible binary output $x \in \{0, 1\}$. The resulting digital signal is then buffered and, if necessary, polarity-adjusted before being fed into the AND gates that compute the third-order interaction products. Under this architecture, the AND gate can serve as the hardware implementation of hyperedge (three-body interactions) during system evolution, which provides a simple and practical solution for realizing probabilistic networks based on hypergraphs (see Supplementary Note 1 and Supplementary Figure 1 for details). The hyperedge computation thus remains fully compatible with conventional digital logic implementation.

### Numerical simulation on various 3-SAT instances

To evaluate the performance of the proposed hypergraph-based direct mapping method in comparison with the logic synthesis approach, a customized 3-SAT instance was analyzed as an initial case study. As shown in **Fig. 4(a)**, this instance is composed of 8 literals and 12 clauses with 28 valid satisfying assignments. This instance presents a slightly complex yet measurable test case. **Fig. 4(b)** presents the merged SUGs that use the conventional approach. This configuration totally comprises 29 vertices. Vertex 29 is clamped to a state of 1 to enforce correct solution constraints. Vertices 1 to 8 serve as output terminals from which the statistics are collected. Vertices 9 to 28 act as 20 auxiliary nodes. These auxiliary nodes are generated during the logic synthesis process, primarily as a result of merging the output of an INOT gate with the input of an IOR gate, as well as from cascading two IOR gates to implement each clause. From the perspective of structural complexity, this network exhibits a high density of connections, consisting of 75 pairwise edges that interconnect throughout the SUG (**Fig. 4c**). In contrast, the hypergraph-based solver derived through direct mapping is depicted in **Fig. 4(d)**. Although they are designed for the same 3-SAT instance, the hypergraph-based solver exhibits a more streamlined architecture. It involves only 8 nodes, each corresponding directly to one of the 8 literals and simultaneously serving as output terminals. The network has significantly fewer edges, totaling just 24 (12 of them are hyperedges), yet effectively captures all necessary relationships between the variables (**Fig. 4e**). The structural differences between the two solver approaches lead to distinct energy landscapes.

Based on the required number of nodes, the solution space of the conventional solver constructed via merged SUGs can be determined, which spans a vast solution space of $2^{28}$ candidate configurations. The energy landscape of this solver is highly complex and chaotic, with energy levels fluctuating dramatically from −43 to 8. The complexity of this energy landscape arises primarily from the differing energy level distributions associated with INOT and IOR gates. Consequently, during logic synthesis, these mismatched energy levels accumulate and stack and finally lead to a rugged landscape characterized by numerous local minima. In contrast, **Fig. 5(a)** shows the energy landscape of the hypergraph-based solver. The energy landscape is significantly simplified, with the solution space reduced to just $2^8$ candidate configurations. Energy levels are discretized into clear, distinct bands, with the ground state $H_{total} = 0$ clearly visible and easily identifiable among the configurations. Compared to the logic synthesis approach, this direct mapping method allows for the direct design of the solver's energy landscape and its contribution is twofold. On one hand, the reduction in the number of nodes significantly lowers the solution space and complexity of the energy landscape in terms of dimensionality. On the other hand, bypassing the stacking of energy levels helps to avoid the emergence of local minimum in the energy landscape, effectively reducing complexity in depth. We then studied the Hamiltonian distribution of these two solvers in more detail. We further analyzed the Hamiltonian distributions of the two solvers in greater detail. As shown in **Fig. 5(b)**, the hypergraph-based design exhibits only a few discrete energy levels within the range of 0-5, corresponding to $N_{EL} = 6$, with a prominently populated ground state. In contrast, the SUG-based solver shows a much richer energy spectrum, with 51 distinct energy levels.

Further simulated experiments comparing the two solvers revealed distinct probability distributions for the given 8-literal and 12-clause 3-SAT problem instance. The two-body logic synthesis SUG-based solver's results, shown in **Fig. 5(c)**, present a highly heterogeneous distribution. Sharp peaks with different heights represent valid solutions that are irregularly dispersed across the configuration space. This pattern suggests that the optimal solutions do not have a theoretically equal probability distribution. Instead, there are numerous local optima in this complex energy landscape. In contrast, the hypergraph-based solver's results, shown in **Fig. 5(d)**, exhibit a more uniform and structured pattern. The 28 optimal solutions, i.e. valid satisfying assignments, exhibit almost equiprobable states, represented as uniformly sized bars. This regular structure implies that the optimal solutions are mapped to the same ground state with $H_{total} = 0$ in the energy landscape.

For the scalability study, we then randomly selected one 3-SAT instance every 100 instances from more challenging SATLIB uf100-430 dataset (instances with 100 literals and 430 clauses). (see Supplementary Note 2 and Supplementary Figure 2 for the uf20-91 SATLIB instances). We first compared the node overhead of the SUG-based and the proposed hypergraph-based methods. As shown in **Fig. 6(a)**, the SUG-based method consistently requires 631 nodes across all instances, whereas the hypergraph-based method uses exactly 100 nodes, strictly equal to the number of variables. The required node count is reduced by ~84%. This substantial difference originates from the fundamentally different construction mechanisms: the SUG-based approach decomposes each clause using INOT and IOR gates, which introduces auxiliary nodes, while the hypergraph-based approach directly maps each clause into a three-body hyperedge term, avoiding logic cascading and node proliferation. In terms of connectivity (**Fig. 6b**), the SUG-based method exhibits an average of ~2,423 pairwise edges across the 10 instances, whereas the hypergraph-based method shows an average of ~1,013 edges (including an average of 430 three-body hyperedges). This introduces an overall reduction of about 58% in edges.

Regarding solving performance, both methods were evaluated using simulated annealing with 5,000 annealing steps, and each data point was obtained from 100 independent runs. As shown in **Fig. 6(c)**, the SUG-based method achieves a success rate of 0% across all tested instances. This indicates that the conventional logic-synthesis-based SUG architecture fails to solve the problem at this scale. In contrast, the hypergraph-based method maintains significantly higher success with an average of ~ 77.6%. This demonstrates that, when the problem size increases to 100 variables, structural complexity differences directly manifest in solution capability. For the average minimum energy identified during the search (**Fig. 6d**), the SUG-based method yields values of 8.74, 10.33, 10.17, 9.76, 8.80, 8.66, 9.06, 8.46, 8.85, and 8.32, with an average of ~9.12. In contrast, the hypergraph-based method achieves an average minimum energy of approximately 0.24, which is close to the theoretical ground state (0), whereas the SUG-based architecture remains trapped at higher energy levels. **Fig. 6(e)** shows two representative energy evolution trajectories for the uf100-01 instance under the two architectures (100 independent runs). As shown, the hypergraph-based design exhibits a faster energy reduction and stable convergence toward the ground state. In contrast, the SUG-based design (**Fig. 6f**) shows slower energy decay and prolonged stagnation at higher energy levels. These results underscore the effectiveness of the hypergraph-based approach with direct mapping in navigating complex energy landscapes and finding optimal solutions efficiently.

## Discussion

This study introduces a hypergraph-based probabilistic computing framework that effectively addresses the limitations of traditional SAT solvers in handling large-scale 3-SAT problems. By employing a direct mapping from logical expressions to hypergraphs, our approach bypasses the unavoidable logic decomposition and synthesis procedures in conventional probabilistic SAT solvers. It significantly reduces the node count and connectivity complexity and achieves the theoretically minimum solution space. Simulation results show that the proposed approach has great benefits in improving the performance of probabilistic 3-SAT solvers. For example, on the uf100-430 benchmark, the proposed framework reduces node count by approximately 84% and edge count by about 58%. Under identical annealing conditions, it achieves an average success rate of ~77.6%, in contrast to the 0% success rate observed for the logic-synthesis-based SUG architecture. Furthermore, the energy landscape of our hypergraph-based solver exhibited fewer local minima and more uniform Hamiltonian distributions, leading to a more robust search for optimal solutions.

While this work establishes a successful proof-of-concept, it is also important to position these results within the broader landscape of mature deterministic SAT solvers. State-of-the-art solvers, including those consistently leading the annual SAT Competitions (e.g., Kissat[42] and CaDiCa[43]), are built upon highly optimized conflict-driven clause learning (CDCL) algorithms and are extremely effective at pruning vast search spaces in large-scale industrial instances. By contrast, the proposed probabilistic hypergraph framework follows a fundamentally different computational paradigm and is not intended as a replacement for such software-based approaches. Scaling this framework to larger and more complex benchmarks introduces several practical challenges. As the number of variables and clauses increases, the density of higher-order interactions grows accordingly, which in turn raises routing complexity and interconnect overhead in hardware implementations. Moreover, p-bit device non-idealities and variability may increasingly influence convergence behavior at scale. Overcoming these limitations will require careful architectural design and continued algorithm–hardware co-optimization. Nevertheless, our findings highlight the effectiveness of the hypergraph-based direct mapping strategy in reducing structural overhead in probabilistic formulations and demonstrate its promise as a complementary direction for future research in probabilistic computing hardware and combinatorial optimization.

## Methods

### BM-Based energy model

The physical mechanism underlying probabilistic 3-SAT solvers is rooted in the Boltzmann Law. Once the system's configuration, i.e., the interconnection relationship, including bias terms, pairwise interaction terms and/or three-body interaction terms, is established, the energy of the system solely depends on the state of the nodes. The steady probability for each state configuration can then be described by the Boltzmann distribution:

$$P(\{x\}) = \frac{e^{-\alpha \cdot H(\{x\})}}{\sum_{\{x\}'} e^{-\alpha \cdot H(\{x\}')}} \tag{6}$$

where $\alpha$ represents a pseudo-temperature parameter indicating the stochasticity level of the system. The encoding of the solution into the ground state is essential in the process of solving the SAT problem, as configurations with the lowest energy are naturally favored during temporal evolution.

### Stochastic State Update Rule and Annealing Mechanism

The proposed solver follows a sequential (asynchronous) stochastic update scheme consistent with Boltzmann-machine dynamics (see Supplementary Note 3 and Supplementary Figure 3 for details). At each annealing step $t$, the p-bits are updated one by one using the most recent states of all other variables. For a given p-bit $x_i \in \{0,1\}$, the effective input signal is obtained from the derivative of the total energy function:

$$I_i = -\frac{\partial H}{\partial x_i} = -\left(h_i + \sum_{j \neq i} J_{ij}\, x_j + \sum_{j<k} W_{ijk}\, x_j x_k\right). \tag{7}$$

The state of $x_i$ is then updated probabilistically according to

$$x_i(t+1) = \begin{cases} 1, & \text{with probability } \sigma(G_0[t] \cdot I_i), \\ 0, & \text{otherwise}, \end{cases} \tag{8}$$

and

$$\sigma(z) = \frac{1}{1+e^{-\alpha z}} \tag{9}$$

where $\sigma(\cdot)$ is the sigmoid activation function, and $G_0[t]$ represents a global annealing scaling factor that modulates the effective interaction strength over iterations. The annealing schedule $G_0[t]$ is gradually increased in a linear manner during the annealing process to balance stochastic exploration at early stages and convergence toward low-energy states at later stages.

## Acknowledgements

This work at the National University of Singapore was supported by FRC-A-8000194-01-00. G.C. L. would also like to thank the financial support from the National Science and Technology Council (NSTC) under grant number NSTC 112-2112-M-A49-047-MY3, the Co-creation Platform of the Industry-Academia Innovation School, NYCU, under the framework of the National Key Fields Industry-University Cooperation and Skilled Personnel Training Act, and the Advanced Semiconductor Technology Research Center from The Featured Areas Research Center Program within the framework of the Higher Education Sprout Project by the Ministry of Education (MOE) in Taiwan.

## Data availability

The generated and processed data are provided in the Supplementary Information/Source Data file. Source data are provided with this paper.

## Code availability

The computer code and problem instances used in this study are available from the corresponding authors upon reasonable request.

## Author contributions

H.Y. and L.G. conceptualized the study and proposed the idea of solving Boolean satisfiability problems with using the hypergraph-based method under the probabilistic computing paradigm. H.Y. designed and conducted simulation experiments with the help of Z.W. H.M. fabricated the p-bit device with help from S.C., L.H., H.Y.C, and W.J. H.Y., G.X, H.T., and L.G., analyzed the simulation and experimental data. H.Y. and H.M. drafted the manuscript with input from all other authors. All authors discussed the results and reviewed the manuscript.

## Corresponding authors

Correspondence and requests for materials should be addressed to L.G. (Email: gcliang@nycu.edu.tw).

## Competing interests

The authors declare no competing financial or non-financial interests.

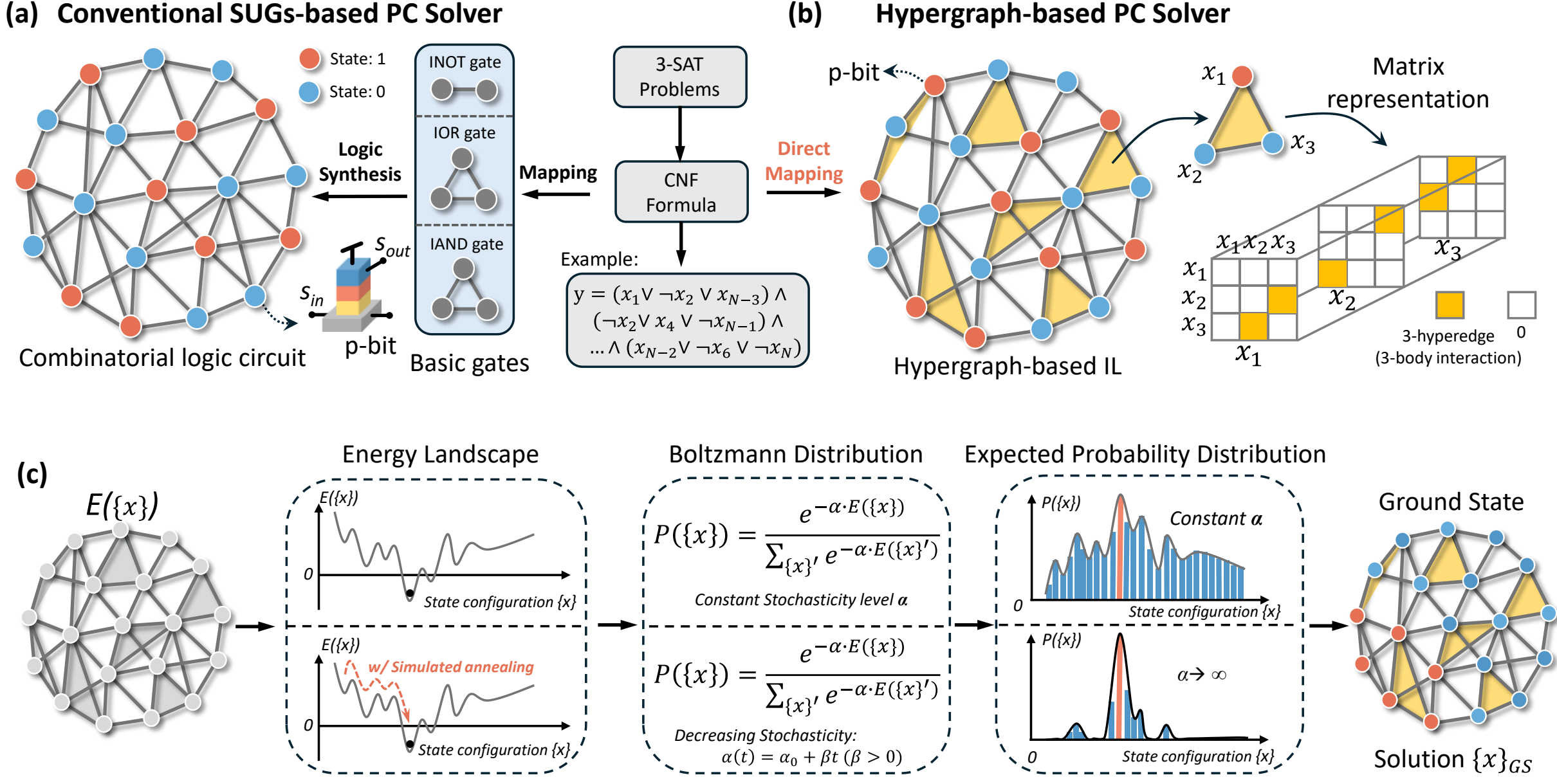

**Fig. 1** Comparative illustration of conventional and proposed hypergraph-based probabilistic computing approaches for solving the 3-SAT problem. (a) Conventional probabilistic SAT solvers constructed from merged SUGs. These solvers typically synthesize basic invertible logic gates, including INOT, IOR, and IAND gates to form a combinatorial probabilistic solver. Vertices/nodes represent binary variables and are implemented with p-bit devices. Pairwise interactions among vertices are depicted by black lines. (b) The proposed solver based on direct mapping from conjunctive normal form (CNF) logical expressions to hypergraph structures. Each hyperedge in yellow-colored region represents higher-order interactions involving multiple vertices, corresponding directly to the clauses within a 3-SAT problem. (c) Underlying physical mechanisms of solving the SAT within probabilistic computing paradigm. The probabilistic dynamics of the proposed system are governed by the Boltzmann distribution, where the probability of a state configuration *{x}* depends on its energy *E({x})* and real-time stochasticity level $\alpha$ of the system. To enhance convergence, simulated annealing is also employed, which gradually sharpens the sampling distribution toward the ground state.

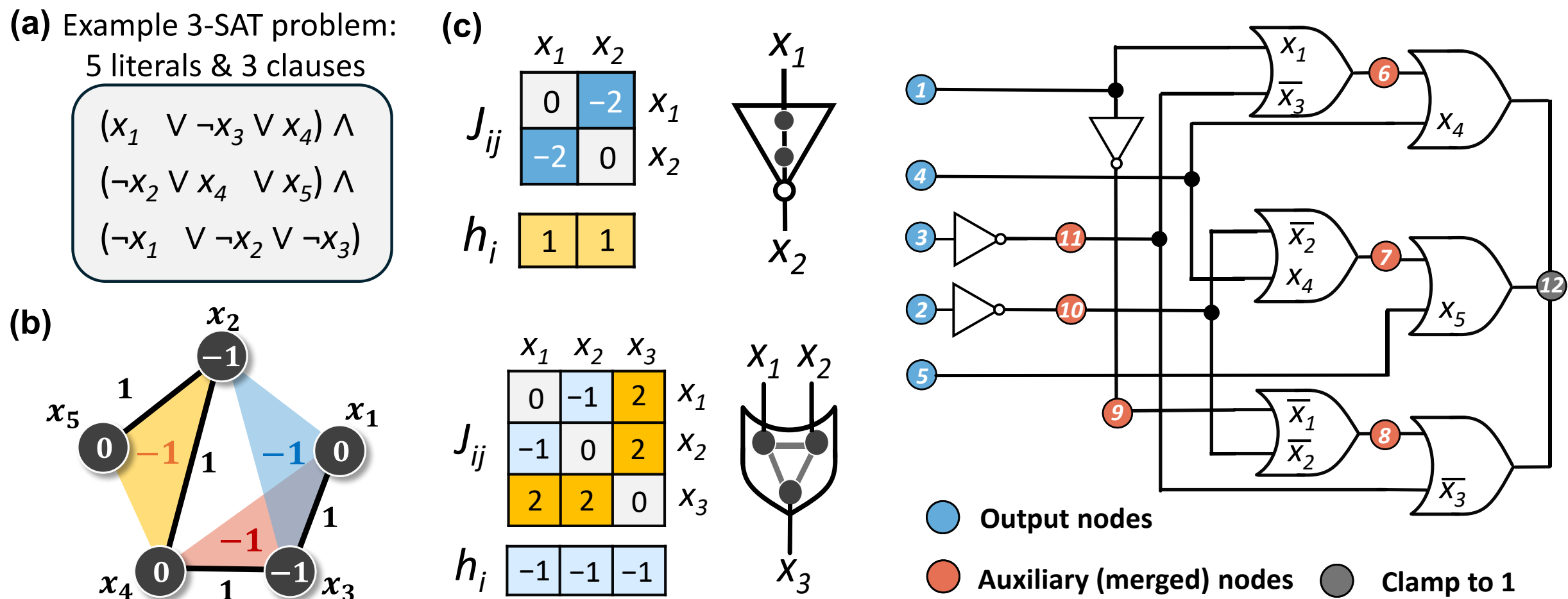

**Fig. 2** (a) Example small-scale 3-SAT problem with 5 literals and 3 clauses. (b) Hypergraph representation of the 3-SAT problem after direct mapping. Hyperedges are illustrated by three colored regions: red, yellow and blue, respectively. Pairwise interactions between vertices are depicted by black lines. The strength of hyperedges and pairwise interactions are indicated by the numbers adjacent to the regions and lines. Local bias term and its strength are represented as white numbers within the vertices. (c) INOT and IOR gates utilized in SUGs-based probabilistic 3-SAT solvers. Each gate is defined by a pairwise interaction matrix $J_{ij}$ and an external bias vector $h_i$. In conventional SUG-based solvers, these two kinds of basic gates undergo logic synthesis and merge shown as the diagram on the right-hand side. In this configuration, the gray node is clamped to logic "1" to enforce correct logical functionality, the blue nodes correspond to observable output variables, and the auxiliary nodes introduced during the merging procedure are highlighted in orange.

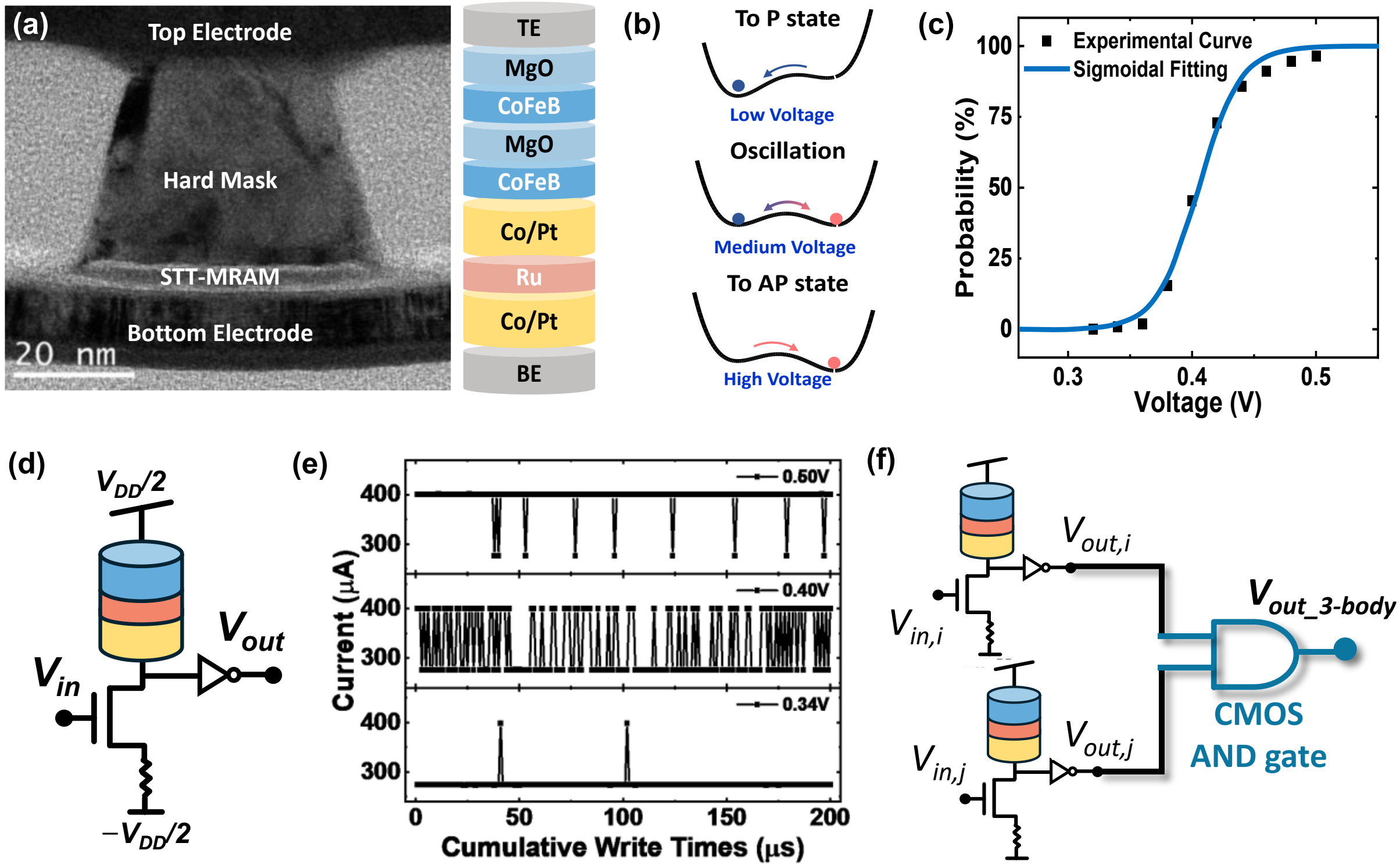

**Fig. 3** (a) Cross-sectional TEM image of a Spin-Transfer Torque Magnetic Random-Access Memory (STT-MRAM) structure. The labeled layers include the Top Electrode (TE), MgO barrier, CoFeB free and fixed layers, Co/Pt multilayers, Ru spacer, and Bottom Electrode (BE). (b) Voltage-dependent switching behavior of the STT-MRAM. (c) Output probability versus voltage characteristics of the magnetic tunnel junction, presenting experimental data (black dots) alongside the sigmoidal fitting curve (blue line). The fitting curve was used in subsequent simulations for solving the 3-SAT problems. (d) Schematic representation of the stochastic STT-biased p-bit device (e) State fluctuation over time under varying input voltages, representing the tunable stochasticity of the tunable p-bit device. (f) Prospective hardware of a conventional CMOS AND gate for implementing three-order interactions in binary output systems.

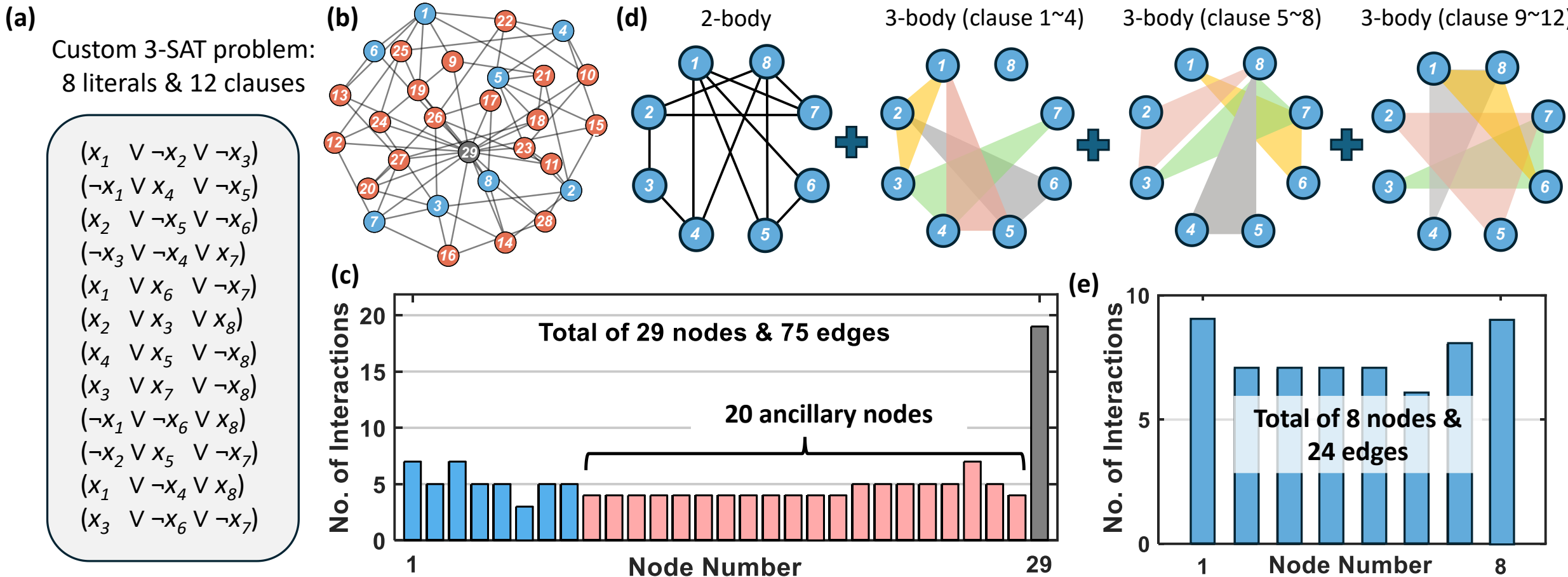

**Fig. 4** (a) Custom-defined 3-SAT problem utilized for evaluating solver performance, with clauses interconnected through logical conjunction. (b) Graphical representation and (c) node-interaction distribution of a conventional 3-SAT solver constructed using the logic synthesis method based on INOT and IOR SUGs. (d) Direct hypergraph-based mapping for the same 3-SAT problem. For better visualization, hyperedges for the 12 clauses are divided into three parts and presented respectively with colored regions enclosing multiple vertices. (e) Node and edges distribution of the 3-SAT solver based on the direct mapping method.

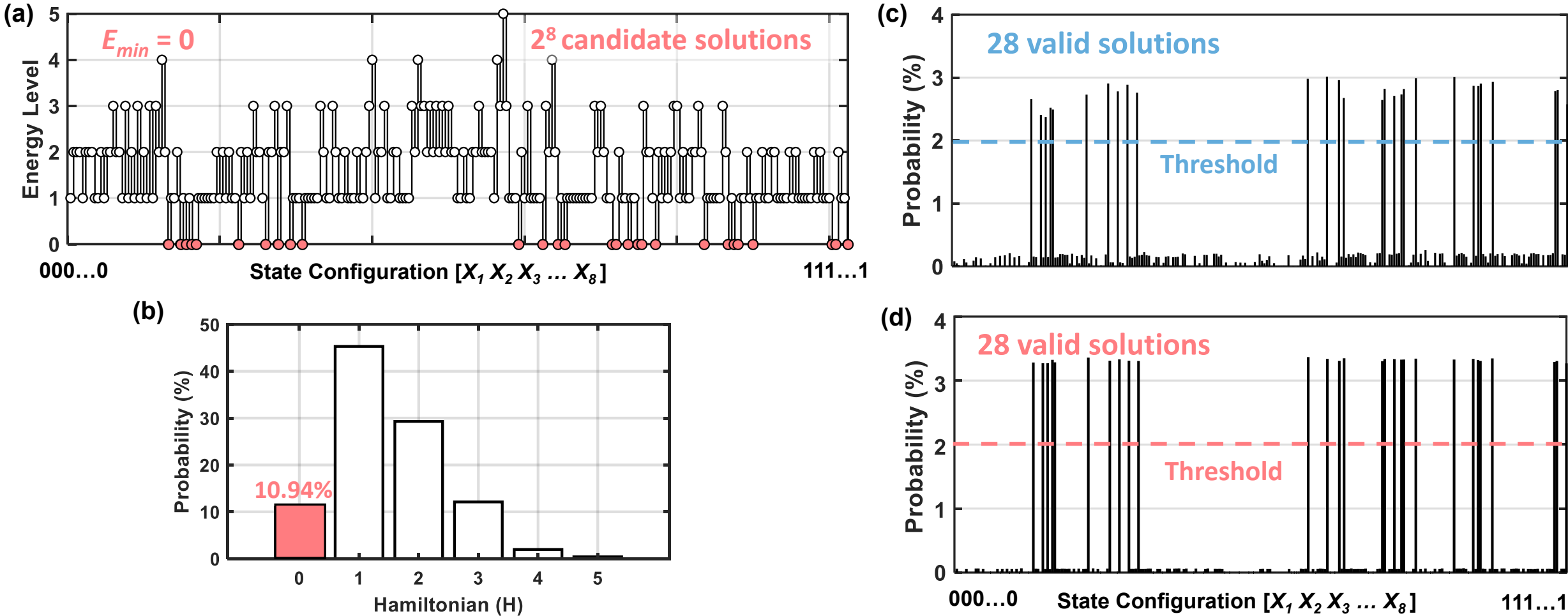

**Fig. 5** Energy landscapes of 3-SAT solvers based on (a) the hypergraph with direct mapping. (b) Hamiltonian distributions of the hypergraph with direct mapping. Probability distributions of the solvers based on (c) the logic synthesis SUG-based approach (29-node) and (d) the hypergraph with direct mapping (8-node).

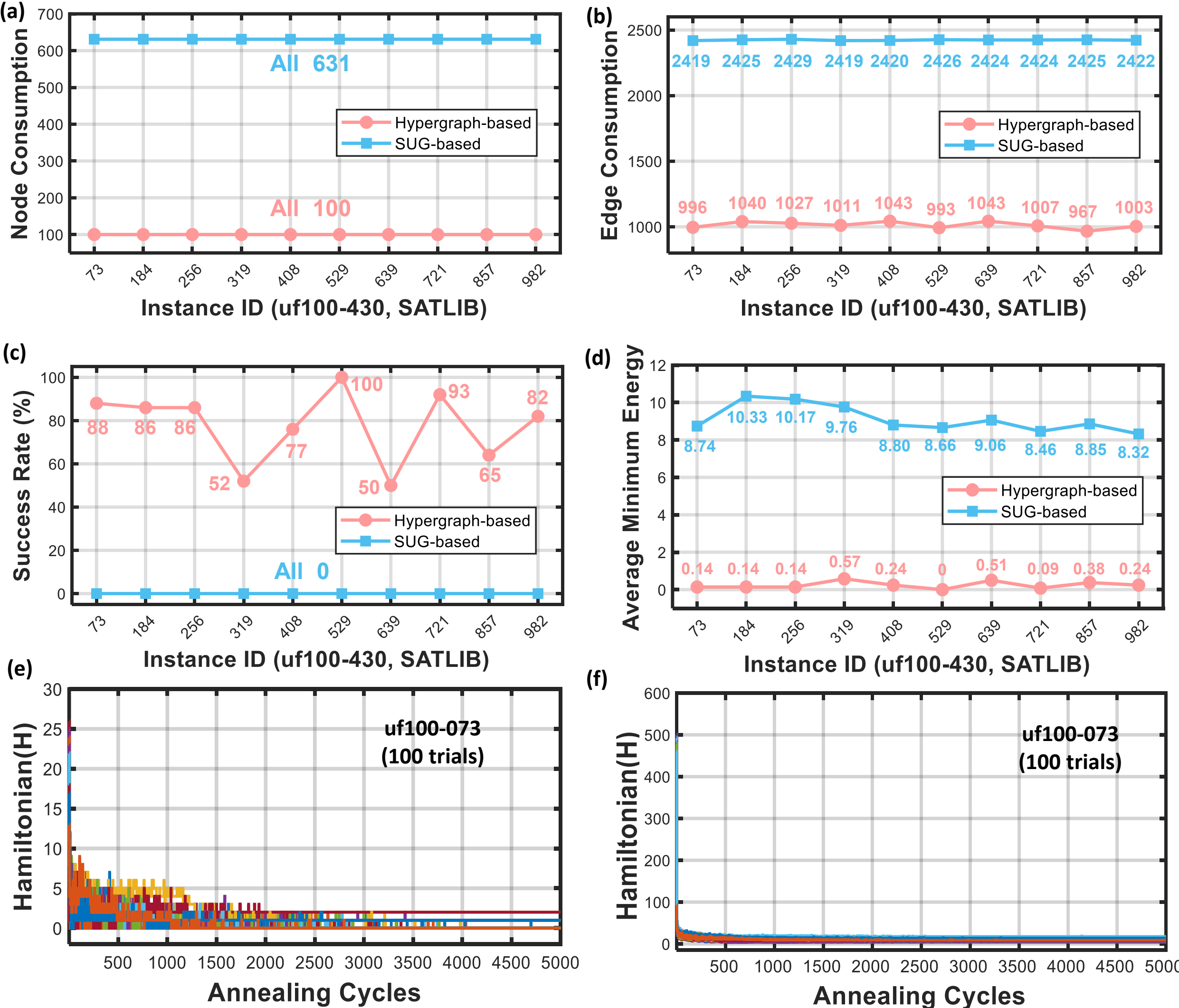

**Fig. 6** Performance comparison between the hypergraph-based and SUG-based designs on the SATLIB uf100-430 benchmark. (a) Node consumption for each instance. (b) Edge consumption for each instance, where for the SUG-based design the count corresponds to the total number of pairwise interaction edges generated by the logic-synthesis-based mapping, and for the hypergraph-based design the count includes both pairwise coupling edges and 3-body hyperedges. (c) Success rate in solving uf100-73, uf100-184, …, and uf100-982. (d) Average minimum energy identified in the search for each instance under identical simulated annealing conditions. (e) Representative energy evolution trajectories of the hypergraph-based design and (f) the SUG-based design for the uf100-73 instance. The evaluation is conducted on 10 randomly selected instances (one sampled every 100 instances from the 1000-instance dataset), with each instance assessed over 100 independent runs using 5000 annealing steps.

# Supplementary Material for

## Solving Boolean Satisfiability Problems Using A Hypergraph-based Probabilistic Computer

Yihan He[1†], Ming-Chun Hong[2,3†],Wanli Zheng[1],Ching Shih[2,3], Hsin-Han Lee[3], Yu-Chen Hsin[3], Jeng-Hua Wei[3], Xiao Gong[1], Tuo-Hung Hou[2,4], and Gengchiau Liang[1,4*]

[1]Department of Electrical and Computer Engineering, National University of Singapore, 117583 Singapore

[2]Department of Electrical Engineering and Institute of Electronics, National Yang-Ming Chiao Tung University, Hsinchu, Taiwan

[3]Electronic and Optoelectronic System Research Laboratories, Industrial Technology Research Institute, Hsinchu, Taiwan

[4]Industry Academia Innovation School, National Yang-Ming Chiao Tung University, Hsinchu, Taiwan

[†]These authors contribute equally; [*]Email: gcliang@nycu.edu.tw

**This PDF file includes:**

Supplementary Note 1 to 3 (incorporating Supplementary Figure 1, Supplementary Figure 2 and Supplementary Figure 3).

**Supplementary Note 1: Hardware Implementation of Hyperedges with Standard CMOS AND Gates**

In the proposed framework, a 3-body interaction associated with a hyperedge $(i, j, k)$is expressed as:

$$H_{ijk} = W_{ijk}\, x_i x_j x_k, \tag{1}$$

where $W_{ijk}$ is the hyperedge coefficient and $x_i \in \{0,1\}$ denotes the binary state of p-bit $i$. Within the Boltzmann-machine formulation, the analog input to each p-bit is determined by the negative derivative of the total energy with respect to its state variable. Taking the derivative of the third-order term with respect to $x_i$ yields:

$$I_i^{(\text{3-body})} = -\frac{\partial\left(W_{ijk} x_i x_j x_k\right)}{\partial x_i} = -W_{ijk} x_j x_k. \tag{2}$$

Similarly, p-bit $j$ receives a contribution $-W_{ijk} x_i x_k$, and p-bit $k$ receives $-W_{ijk} x_i x_j$. Therefore, a single hyperedge naturally gives rise to three symmetric feedback paths, one toward each participating p-bit, fully consistent with Boltzmann dynamics.

At the circuit level (Supplementary Figure 1), the term $x_j x_k$ is realized using local CMOS logic gates that generate the binary product of the corresponding p-bit outputs. Specifically, two-input AND gates compute the pairwise product (e.g., $x_j \wedge x_k = x_j x_k$, since $x \in \{0,1\}$). The resulting binary signal is subsequently weighted by the programmable coefficient $-W_{ijk}$ and injected into the analog summation node of the target p-bit. Importantly, the logic gate itself only produces the binary product, while the amplitude and sign of the interaction are controlled by the programmable weight stage. In this manner, third-order interactions are incorporated into the same physical accumulation pathway used for pairwise couplings and bias terms.

Consequently, the total analog input of each p-bit follows

$$I_i = -\left(W_{ijk} \cdot (x_j x_k) + J_{ij} x_j + J_{ik} x_k + h_i\right), \tag{3}$$

$$I_j = -\left(W_{ijk} \cdot (x_i x_k) + J_{ji} x_i + J_{ik} x_k + h_j\right), \tag{4}$$

$$I_k = -\left(W_{ijk} \cdot (x_i x_j) + J_{ki} x_i + J_{kj} x_j + h_k\right). \tag{5}$$

The framework naturally generalizes to weighted clauses. If a clause carries an explicit weight, as in weighted SAT formulations, this weight can be directly absorbed into the corresponding hyperedge coefficient $-W_{ijk}$. In a hardware realization, such coefficients can be implemented digitally using programmable multipliers (e.g., lookup-table-based multipliers on an FPGA platform) or analog weight elements, which scale the binary product before injection into the p-bit input.

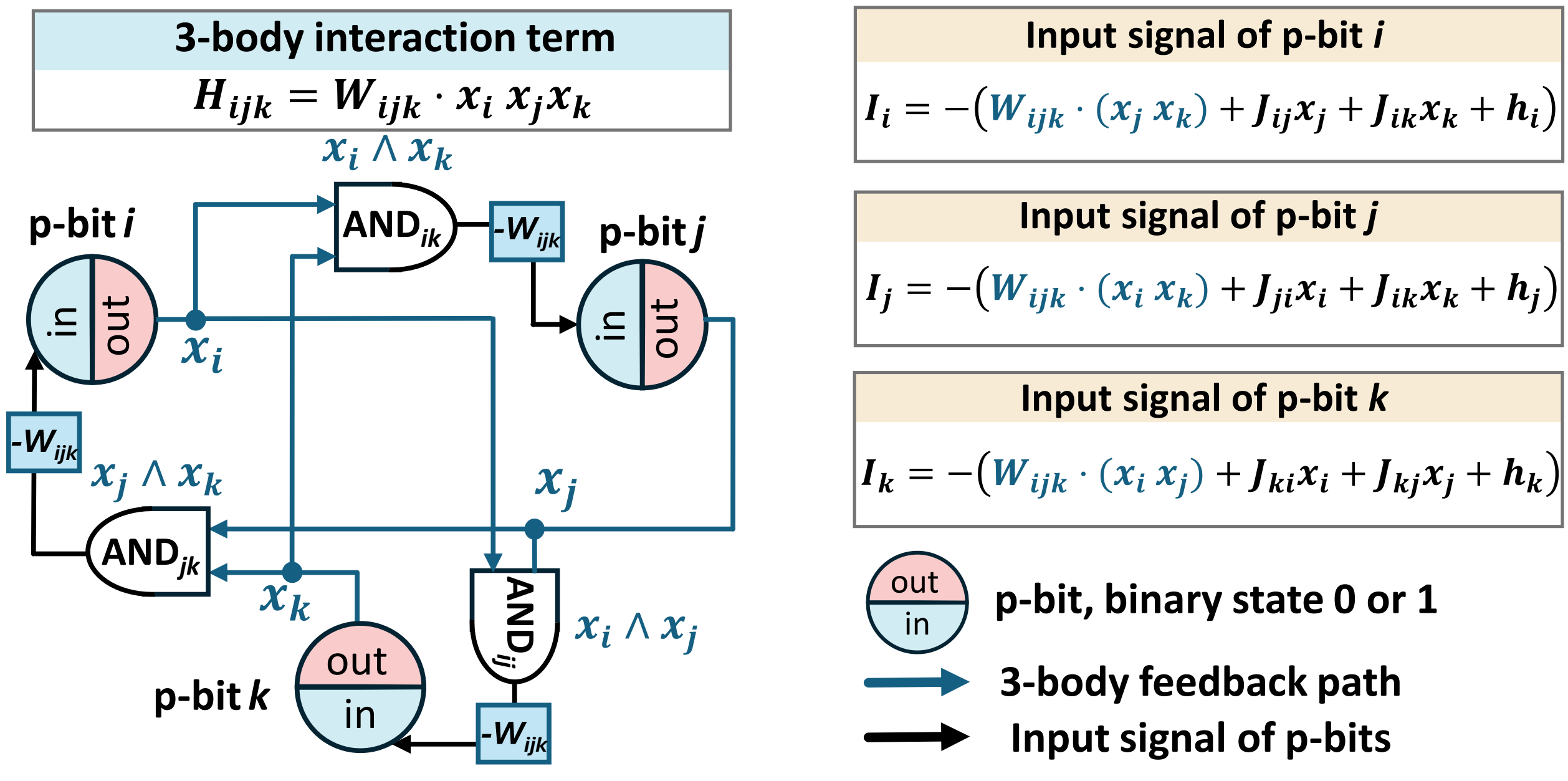

**Supplementary Figure 1** Circuit-level schematic for realizing a three-body hyperedge with a symmetric feedback topology. A single hyperedge connecting p-bits $i$, $j$, and $k$ is constructed using three 2-input CMOS AND gates arranged in a closed-loop feedback configuration. Each AND gate computes one pairwise state product and routes the weighted output back to the input node of the remaining p-bit, forming three reciprocal feedback paths. The weighting element $-W_{ijk}$ scales the contribution of each pairwise product before it is injected into the analog input of the corresponding stochastic unit.

**Supplementary Note 2: Extended Evaluation on STALIB uf20-91 Instances**

To further evaluate the performance of the proposed hypergraph-based framework, additional simulations were conducted on benchmark instances from the SATLIB uf20-91 dataset. Ten representative 3-SAT instances were selected by sampling one instance every 100 entries from the dataset of 1000 problems. For each instance, both the SUG-based and hypergraph-based architectures were evaluated under an identical annealing schedule of 2500 steps. All reported metrics were obtained from 100 independent runs with random initializations to ensure statistical robustness.

In the revised SUG-based implementation, the circuit structure follows a standard invertible logic synthesis framework using conventional INOT and IOR gates. Under this construction, the uf20-91 SUG-based solver contains 132 nodes, while the hypergraph-based solver requires only 100 nodes without auxiliary variables. This structural difference reflects the distinct mapping strategies adopted by the two approaches.

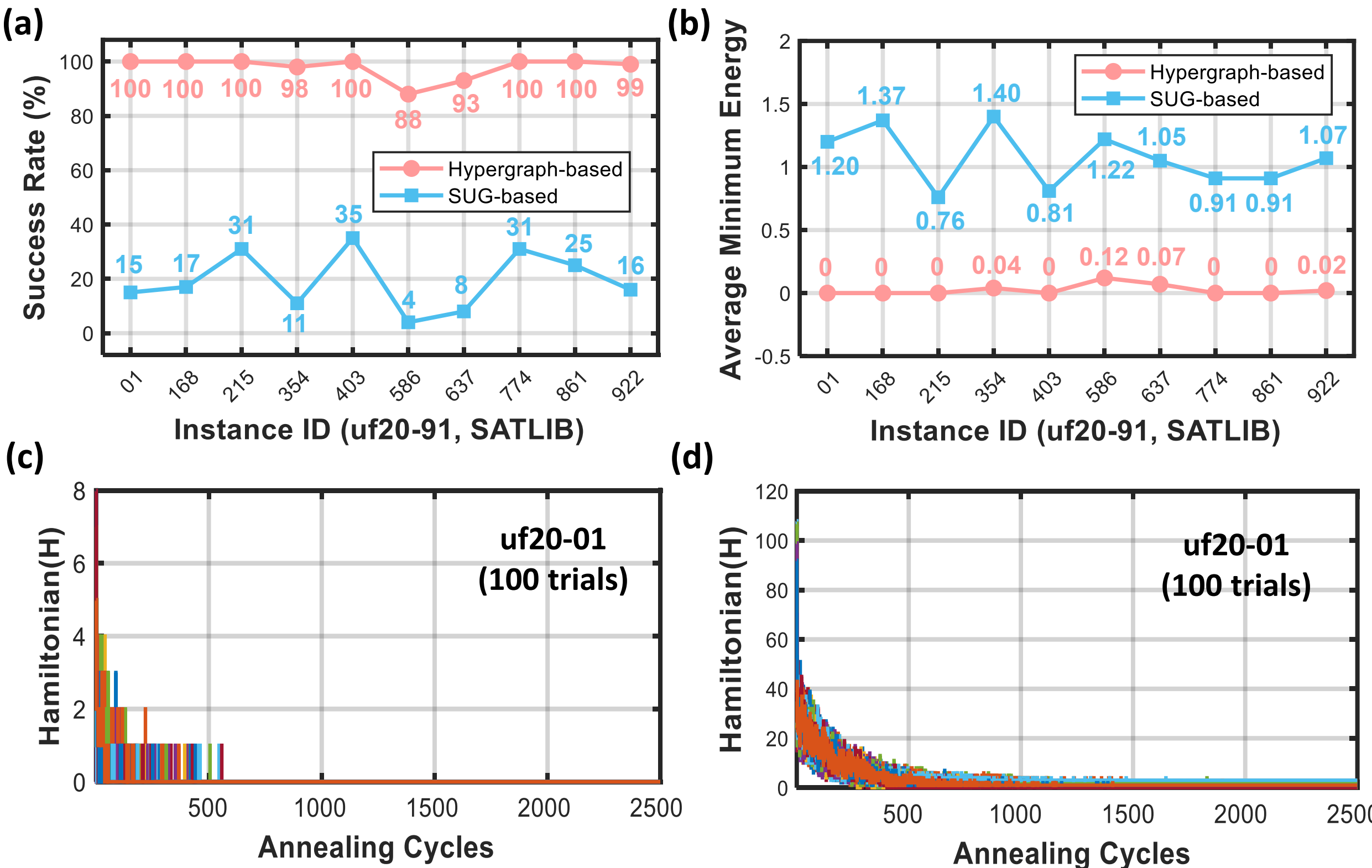

**Supplementary Figure 2** Performance comparison between the hypergraph-based and SUG-based architectures on the SATLIB uf20-91 benchmark. (a) Success rate in solving uf20-01, uf 20-168, …, and uf20-922. (b) Average minimum energy identified in the search for each instance under identical simulated annealing conditions. (c) Representative energy evolution trajectories of the hypergraph-based design and (d) the SUG-based design for the uf20-01 instance. The evaluation is conducted on 10 randomly selected instances (one sampled every 100 instances from the 1000-instance dataset), with each instance assessed over 100 independent runs using 2500 annealing steps.

Across the ten benchmark instances, the hypergraph-based architecture achieved an average success rate of approximately 97.8%, whereas the SUG-based architecture achieved an average success rate of approximately 19.3% (Supplementary Figure 2a). The hypergraph-based method also attained a lower average minimum energy (~0.025), with several instances consistently reaching the theoretical ground state (0), while the SUG-based method exhibited an average minimum energy of approximately 1.07 (Supplementary Figure 2b). Variations in success rate across different instances were observed to correlate with the number of satisfying assignments associated with each problem. Instances with fewer satisfying solutions naturally exhibit lower success probabilities due to the reduced fraction of ground states within the configuration space. This trend was consistently observed across both architectures.

Energy evolution trajectories for the representative instance uf20-01 further highlight the difference in convergence behavior. The hypergraph-based design demonstrates faster energy reduction and stable convergence toward low-energy states (Supplementary Figure 2c), whereas the SUG-based design shows slower energy decay and more frequent stagnation at higher energy levels (Supplementary Figure 2d).

These extended simulations provide additional evidence of the structural efficiency and convergence advantages offered by the hypergraph-based direct mapping strategy.

**Supplementary Note 3: Stochastic State Update Rule and Annealing Mechanism**

The proposed solver operates by performing stochastic energy minimization on a hypergraph-encoded formulation of the 3-SAT problem. The total energy of the system is defined as

$$H_{\text{total}} = \sum_{i<j<k} W_{ijk}\, x_i x_j x_k + \sum_{i<j} J_{ij}\, x_i x_j + \sum_{i} h_i\, x_i + C, \tag{6}$$

where $W_{ijk}$, $J_{ij}$, and $h_i$ denote the coefficients associated with third-order hyperedges, pairwise interactions, and local bias terms, respectively, and $x_i \in \{0,1\}$ represents the binary state of p-bit $i$. The constant term $C$ does not affect the dynamics and is omitted in the update procedure.

For each p-bit, the effective input is obtained from the negative derivative of the energy function with respect to its state variable:

$$I_i = -\frac{\partial H_{\text{total}}}{\partial x_i} = -\left(h_i + \sum\nolimits_{i<j} J_{ij}\, x_j + \sum\nolimits_{j<k} W_{ijk}\, x_j x_k\right). \tag{7}$$

This quantity represents the accumulated interaction acting on p-bit $i$, including contributions from bias, pairwise couplings, and higher-order hyperedges. In a hardware realization, this corresponds to the analog summation of weighted interaction signals at the p-bit input node.

The solver follows a sequential (asynchronous) stochastic update scheme consistent with Boltzmann-machine dynamics. At each annealing step $t$, the p-bits are updated one by one using the most recent states of all other variables. The effective input is modulated by a global annealing scaling factor $G_0[t]$,

$$I_i^{(\text{eff})} = G_0[t] \cdot I_i, \tag{8}$$

and the state is updated probabilistically according to

$$x_i(t+1) = \begin{cases} 1, & \text{with probability } \sigma\left(I_i^{(\text{eff})}\right), \\ 0, & \text{otherwise,} \end{cases} \tag{9}$$

where the sigmoid function is defined as

$$\sigma(z) = \frac{1}{1 + e^{-\alpha z}}. \tag{10}$$

Here, $\alpha$ characterizes the intrinsic stochastic response of the p-bit device and determines the slope of the transfer curve. The global factor $G_0[t]$ therefore controls the effective interaction strength at the network level, while $\alpha$ reflects the inherent device-level randomness. Together, they determine the effective "temperature" of the system.

As the problem size increases, the number of interaction terms contributing to each p-bit grows, leading to a larger dynamic range of the accumulated input. To maintain operation within the effective region of the sigmoid response and avoid saturation, the initial magnitude of $G_0[t]$ is selected according to the instance scale. During optimization, $G_0[t]$ follows a linear annealing schedule, typically increasing over iterations to balance stochastic exploration in early stages with convergence toward low-energy configurations in later stages. Uniform scaling through $G_0[t]$ preserves the relative structure of the energy landscape and does not alter the location of global minima, while enabling controlled modulation of exploration and convergence.

For completeness, the full pseudocode summarizing the update procedure is provided in Supplementary Figure 3.

**Pseudocode: Sequential p-bit Update with 2-body and 3-body Couplings**

- **Inputs:**
  Pairwise coupling matrix $J \in \mathbb{R}^{N\times N}$,
  bias vector $h \in \mathbb{R}^{N}$,
  hyperedge set $\varphi$ where each $(i, j, k, W_{ijk})$ defines a 3-body interaction,
  annealing schedule $G_0[t]$, total steps $T$.

---

- **Initialize**
  Randomly initialize $x \in \{0, 1\}^N$
- **For** $t = 1$ to $T$ **do**
  - **For** $i = 1$ to $N$ ***(sequential / asynchronous update)***
    - Compute effective input analog signal:
      $$I_i \leftarrow -\left(h_i + \sum_{i<j} J_{ij}x_j + \sum_{i<j<k} W_{ijk}x_j \wedge x_k\right)$$
      where $\varphi(i)$ denotes all hyperedges involving node $i$.
    - Perform stochastic p-bit update:
      $$x_i \sim \text{Bernoulli}\big(\sigma(G_0[t] \cdot I_i)\big)$$
      with
      $$\sigma(z) = \frac{1}{1 + e^{-\alpha\cdot z}}$$
      where $\alpha$ represents the pseudo-temperature parameter.
  - Evaluate system energy
    $H(x)$ according to Eq.(4)
    for monitoring or success detection.
- **End For**

**Supplementary Figure 3** Pseudocode of the sequential Boltzmann update procedure used in the hypergraph-based probabilistic 3-SAT solver.